\documentclass[10pt,conference]{IEEEtran}
\IEEEoverridecommandlockouts

\usepackage{cite}
\usepackage{amsmath,amssymb,amsfonts}

\usepackage{algpseudocode}
\usepackage[ruled, linesnumbered, vlined]{algorithm2e} 

\usepackage{graphicx}
\usepackage{balance,amsthm}
\usepackage{acronym}
\usepackage{textcomp}
\usepackage{xcolor}
\graphicspath{ {./images/} }
\usepackage{booktabs}
\usepackage{lscape}
\usepackage{adjustbox}
\usepackage[left=1.62cm,right=1.62cm,top=0.8in]{geometry}
\usepackage{lipsum} 
\usepackage{flafter} 
\usepackage{color,soul} 
\usepackage{tabularx,booktabs} 
\newcolumntype{C}{>{\centering\arraybackslash}X} 
\usepackage[bookmarks=false,hidelinks,draft=false]{hyperref} 
\usepackage{scalerel} 
\usepackage{tikz} 
\usepackage{pgfplots}
\pgfplotsset{compat=newest}
\usetikzlibrary{plotmarks}
\usetikzlibrary{arrows.meta}\usepgfplotslibrary{patchplots}

\renewcommand{\baselinestretch}{1.01}
\def\BibTeX{{\rm B\kern-.05em{\sc i\kern-.025em b}\kern-.08em T\kern-.1667em\lower.7ex\hbox{E}\kern-.125emX}}
\usepackage{stfloats}

\usetikzlibrary{svg.path}
\definecolor{orcidlogocol}{HTML}{A6CE39}
\tikzset{
	orcidlogo/.pic={
		\fill[orcidlogocol] svg{M256,128c0,70.7-57.3,128-128,128C57.3,256,0,198.7,0,128C0,57.3,57.3,0,128,0C198.7,0,256,57.3,256,128z};
		\fill[white] svg{M86.3,186.2H70.9V79.1h15.4v48.4V186.2z}
		svg{M108.9,79.1h41.6c39.6,0,57,28.3,57,53.6c0,27.5-21.5,53.6-56.8,53.6h-41.8V79.1z M124.3,172.4h24.5c34.9,0,42.9-26.5,42.9-39.7c0-21.5-13.7-39.7-43.7-39.7h-23.7V172.4z}
		svg{M88.7,56.8c0,5.5-4.5,10.1-10.1,10.1c-5.6,0-10.1-4.6-10.1-10.1c0-5.6,4.5-10.1,10.1-10.1C84.2,46.7,88.7,51.3,88.7,56.8z};
	}
}
\newcommand{\orcidicon}[1]{\href{https://orcid.org/#1}{\mbox{\scalerel*{
				\begin{tikzpicture}[yscale=-1,transform shape]
				\pic{orcidlogo};
				\end{tikzpicture}
			}{|}}}}

\newcommand{\ignore}[1]{}

\newcommand{\linebreakand}{%
\end{@IEEEauthorhalign}
\hfill\mbox{}\par
\mbox{}\hfill\hspace*{-1cm}\begin{@IEEEauthorhalign} 
}

\newtheorem*{remark}{Remark}

\allowdisplaybreaks
\acrodef{RIS}{reconfigurable intelligent surface}
\acrodef{SNR}{signal-to-noise ratio}
\acrodef{ISAC}{integrated sensing and communication}
\acrodef{ISLAC}{integrated sensing, localization, and communication}
\acrodef{LoS}{line-of-sight}
\acrodef{NLoS}{non-line-of-sight}
\acrodef{AoA}{angle-of-arrival}
\acrodef{AoD}{angle-of-departure}
\acrodef{ToA}{time-of-arrival}
\acrodef{UE}{user equipment}
\acrodef{NF}{noise figure}
\acrodef{PSD}{power spectral density}
\acrodef{BS}{base station}
\acrodef{MCRB}{misspecified Cram\'{e}r-Rao bound}
\acrodef{CRB}{Cram\'{e}r-Rao bound}
\acrodef{LB}{lower bound}
\acrodef{ML}{maximum-likelihood}
\acrodef{MML}{mismatched maximum-likelihood}
\acrodef{DL}{downlink}
\acrodef{UL}{uplink}
\acrodef{MIMO}{multiple-input multiple-output}
\acrodef{MISO}{multiple-input single-output}
\acrodef{SISO}{single-input single-output}
\acrodef{SIP}{shift invariance property}
\acrodef{FIM}{Fisher information matrix}
\acrodef{RMSE}{root mean-squared error}
\acrodef{AWGN}{additive white Gaussian noise}
\acrodef{ADMM}{alternating direction method of multipliers}
\acrodef{LS}{least-squares}
\acrodef{SOC}{second-order cone}
\acrodef{CFO}{carrier frequency offset}
\acrodef{GLRT}{generalized likelihood ratio test}
\acrodef{FSPL}{free space path loss}
\acrodef{TDoA}{time-difference-of-arrival}
\acrodef{MPC}{multi-path components}
\acrodef{NB}{narrowband}
\acrodef{WB}{wideband}
\acrodef{TDM}{time-division multiplexing}
\acrodef{NTN}{non-terrestrial network}
\acrodef{KPI}{key performance indicator}
\acrodef{KVI}{key value indicator}
\acrodef{SDG}{sustainable development goal}
\acrodef{LEO}{low earth orbit}
\acrodef{OFDM}{orthogonal frequency division multiplexing}
\acrodef{LCS}{local coordinate system}
\acrodef{GCS}{global coordinate system}
\acrodef{CP}{cyclic prefix}
\acrodef{BF}{beamforming}
\acrodef{CRLB}{Cramér-Rao lower bound}
\acrodef{MLE}{maximum-likelihood estimator}
\acrodef{URA}{uniform rectangular array}
\acrodef{FDoA}{frequency-difference-of-arrival}
\begin{document}
\bstctlcite{IEEEexample:BSTcontrol}
\title{6G RIS-aided Single-LEO Localization with Slow and Fast Doppler Effects}
\author{ 
Sharief Saleh\IEEEauthorrefmark{1},
Musa Furkan Keskin\IEEEauthorrefmark{1},
Basuki Priyanto\IEEEauthorrefmark{2},
Martin Beale\IEEEauthorrefmark{2},\\
Pinjun Zheng\IEEEauthorrefmark{3},
Tareq Y. Al-Naffouri\IEEEauthorrefmark{3},
Gonzalo Seco-Granados\IEEEauthorrefmark{4},
Henk Wymeersch\IEEEauthorrefmark{1}\\
\IEEEauthorblockA{
\IEEEauthorrefmark{1}Chalmers University of Technology, Gothenburg, Sweden,
\IEEEauthorrefmark{2}Sony Europe, Sweden,\\
\IEEEauthorrefmark{3}King Abdullah University of Science and Technology, Thuwal, KSA,
\IEEEauthorrefmark{4}Universitat Autonoma de Barcelona, Barcelona, Spain\\
email: sharief@chalmers.se, furkan@chalmers.se, gonzalo.seco@uab.cat, henkw@chalmers.se}

\vspace{-4mm}
\thanks{This work is supported by the European Commission through the Horizon Europe/JU SNS project Hexa-X-II (Grant Agreement no. 101095759), the Swedish Research Council grant 2022-03007, the Spanish R+D project PID2020-118984GB-I00, and the King Abdullah University of Science and Technology (KAUST) Office of Sponsored Research (OSR) under Award ORA-CRG2021-4695.}
}

\maketitle

\begin{abstract}	
6G networks aim to enable applications like autonomous driving by providing complementary localization services through key technologies such as non-terrestrial networks (NTNs) with low Earth orbit (LEO) satellites and reconfigurable intelligent surfaces (RIS). Prior research in 6G localization using single LEO, multi-LEO, and multi-LEO multi-RIS setups has limitations: single LEO lacks the required accuracy, while multi-LEO/RIS setups demand many visible satellites and RISs, which is not always feasible in practice. This paper explores the novel problem of localization with a single LEO satellite and a single RIS, bridging these research areas. We present a comprehensive signal model accounting for user carrier frequency offset (CFO), clock bias, and fast and slow Doppler effects. Additionally, we derive a low-complexity estimator that achieves theoretical bounds at high signal-to-noise ratios (SNR). Our results demonstrate the feasibility and accuracy of RIS-aided single-LEO localization in 6G networks and highlight potential research directions.
\end{abstract}

\section{Introduction}
\Acp{NTN} utilize space- and airborne platforms and are expected to play a crucial role in social sustainability, enhancing network resilience, and providing continuous connectivity in scenarios where terrestrial infrastructure may be inadequate or non-existent \cite{azari_evolution_2}. In particular, the use of \ac{LEO} satellites has the potential to provide global 6G integrated communication and sensing (e.g., positioning, imaging) services \cite{luo2024leo}, complementing existing terrestrial communication systems and space-based positioning systems (e.g., GPS) \cite{sallouha2024ground}. 

Focusing on the important application of user localization, \ac{LEO}-based \ac{NTN} has been studied in various contexts, benefiting from the high satellite mobility, high-power signals, and low latencies \cite{dureppagari2023ntn,you2024integrated}.
Existing studies can be divided into multi-\ac{LEO} localization  \cite{emara_positioning_2,chandrika2023spin,singh2022opportunistic,khalife2021first} or single-\ac{LEO} localization \cite{dureppagari2023ntn,you2024integrated,Wang2020}.
\begin{figure}
    \centering
    \includegraphics[trim=0pt 90pt 425pt 0pt,clip,width=0.8\columnwidth]{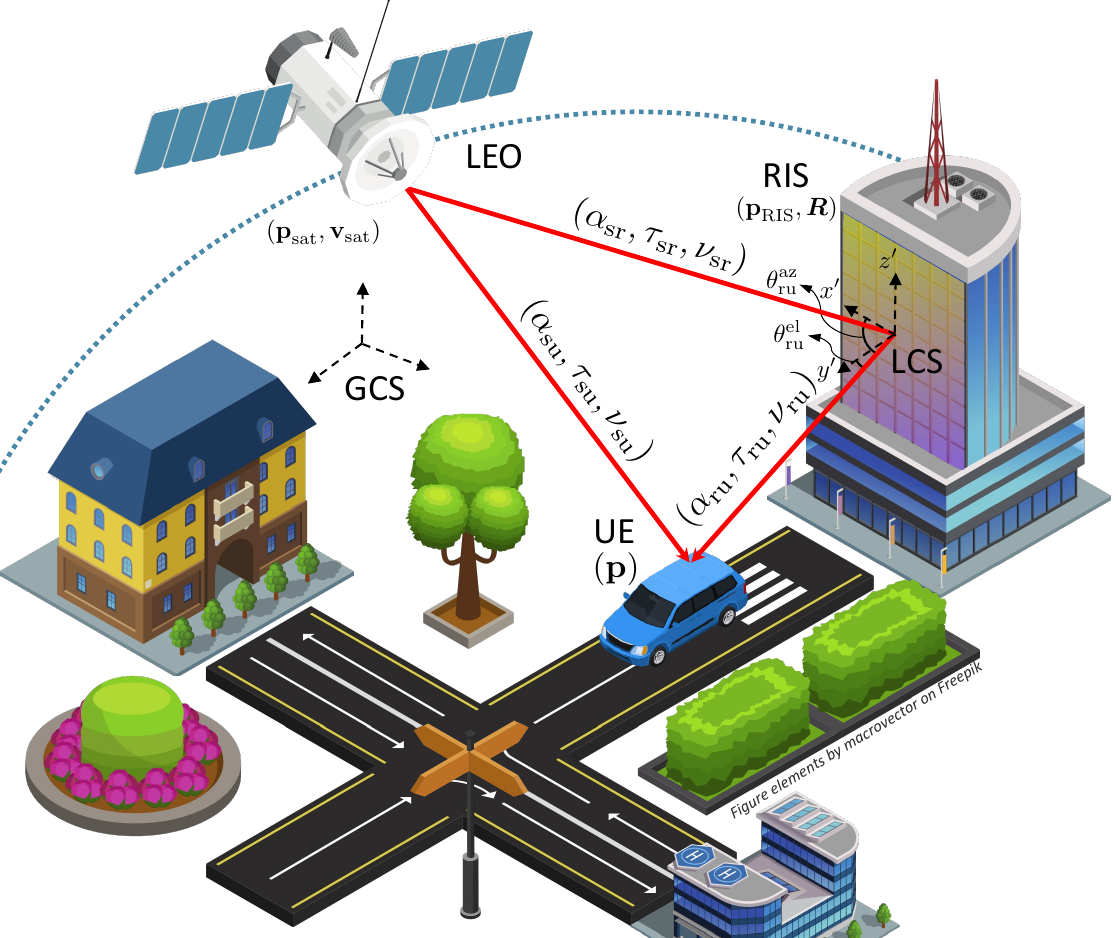}
    \caption{A 6G localization scenario with a single-LEO satellite aided by a single-RIS in an urban setting.}
    \label{fig:SysModel}
\end{figure}
In the former category, \cite{emara_positioning_2} studied a multi-\ac{LEO} system with  3GPP numerology, while \cite{chandrika2023spin} focused on joint localization and time-frequency synchronization in \ac{NTN}-IoT applications. Opportunistic multi-\ac{LEO} localization using \ac{NTN} was proposed in \cite{singh2022opportunistic}, while 
\cite{khalife2021first} demonstrated the practical feasibility of positioning without detailed knowledge of the transmitted signal. In the latter category,  \cite{you2024integrated} considered joint localization and communication from a single \ac{LEO} satellite of a time-synchronized user, relying on \ac{AoD} and \ac{ToA} measurements. Note here that small \ac{AoD} measurement errors will lead to high positioning errors at such a long distance. On the other hand, \cite{Wang2020} utilized \ac{TDoA} and \ac{FDoA} of signals sent at two-time instances, i.e., capitalizing on the satellite's high mobility, to localize the user. Such a method requires minimal \ac{UE} mobility between the two transmissions, ideally a static \ac{UE}, for it to work. 
An alternative way to solve the 6G \ac{LEO} positioning problem without needing terrestrial \acp{BS}, unreliable \ac{LEO}-\ac{AoD} measurements, or long inter-transmission times is to rely on \acp{RIS}, as mentioned in \cite{azari_evolution_2,luo2024leo,sallouha2024ground} and elaborated in \cite{zheng2023leo,wang2024beamforming,zheng2024leo}. For instance, \cite{zheng2023leo} showed that so-called STAR-\ac{RIS} can enhance single-\ac{LEO} indoor localization coverage through theoretical error-bound analysis. An \ac{RIS} beamforming technique to optimize the positioning \ac{CRLB} of a single satellite-\ac{RIS}-\ac{UE} problem was proposed in \cite{wang2024beamforming}. The proposed method requires knowledge about the \ac{RIS}-\ac{UE} \ac{AoD}, which cannot be estimated due to the fixed \ac{RIS} configuration over time. To solve that, they assume rough prior knowledge about the UE's position, from which a rough AOD estimate can be attained. In\cite{zheng2024leo}, a Riemannian manifold-based approach was proposed to solve a 9D \ac{UE} tracking problem (3D position, velocity, and orientation) in a multi-\ac{LEO} multi-\ac{RIS} setup. This makes \cite{zheng2024leo} the only work, to date, to attempt to solve the \ac{RIS}-aided \ac{LEO} localization estimation problem, albeit with multiple satellites and \acp{RIS} in view, which imposes stringent constraints in the constellation deployment. 

In this paper, inspired by \cite{keykhosravi_ris-enabled_2022},  
we address the problem of single-\ac{LEO} user localization with the aid of a single \ac{RIS} while considering the impact of time-varying \ac{RIS} configurations on the capability to estimate the \ac{AoD} and to separate the \ac{RIS} path from the direct path. The contributions of this work are two-fold: (i) we present a more complete and realistic signal model than that used in \cite{zheng2023leo,wang2024beamforming,zheng2024leo}, accounting for both the user \ac{CFO} and clock bias, as well as slow-time and fast-time Doppler effects; and (ii) we derive a practical multi-stage estimator that harnesses the problem structure to circumvent the coupling between the AoD and the CFO/Doppler in time-domain. 

\subsubsection*{Notation} 
We use bold for (column) vectors (e.g., $\boldsymbol{x}$) and bold uppercase for matrices (e.g., $\boldsymbol{X}$). Transpose is denoted as $\boldsymbol{X}^\top$, Hermitian as $\boldsymbol{X}^\mathsf{H}$, and complex conjugate as $\boldsymbol{X}^*$. The entry on row $k$, column $l$ of matrix $\boldsymbol{X}$ is denoted by $X_{k,l}$.

\section{System Model}
We consider a single \ac{LEO} satellite with a single directional antenna, a single \ac{UE} with an omnidirectional antenna, and a single \ac{RIS} in the vicinity of the \ac{UE}, as shown in Fig. \ref{fig:SysModel}. We assume that the \ac{UE} is static and has an unknown location $\boldsymbol{p}\in\mathbb{R}^3$, whereas the \ac{RIS} has a known location  $\boldsymbol{p}_\text{RIS}\in \mathbb{R}^3$, and a known orientation described by $\boldsymbol{R}\in\text{SO}(3)$, the rotation from the \ac{LCS} of the RIS to the \ac{GCS} \cite{sallouha2024ground}. During the measurement time, the satellite has known location $\boldsymbol{p}_\text{sat}(t)=\boldsymbol{p}_\text{sat}(0)+t\boldsymbol{v}_\text{sat}$ and velocity $\boldsymbol{v}_\text{sat}$. The clocks of the satellite and the \ac{UE} are assumed to have an unknown time offset  $\delta$, and an unknown \ac{CFO} $\delta_f$. 
The \ac{RIS} is equipped with a \ac{URA} of $N=N_x\times N_z$ elements, where $N_x$ and $N_z$ are the number of the \ac{RIS} elements placed along its local x and z-axes, respectively. The $n^{\text{th}}$ element of the \ac{RIS} is located at 
$\boldsymbol{p}_n$ in the \ac{LCS} of the \ac{RIS}. The \ac{RIS} elements are spaced by $\lambda/2$, where $\lambda=c/f_c$ is the wavelength, $f_c$ is the carrier frequency, and $c$ is the speed of light. The \ac{RIS} phase configuration is denoted by $\boldsymbol{\omega}=[e^{\jmath \omega_{1}},\dots,e^{\jmath \omega_{N}}]^{\top}$, where $\omega_{n}=-2\pi f \tau_{n}$ is the phase shift induced by the $n^{\text{th}}$ element's delay  $\tau_{n}\in[0,1/f_c]$ \cite{bjornson2022reconfigurable}.

\subsection{Signal and Channel Models}
\subsubsection{Transmit Signal Model}
The satellite emits \ac{OFDM} signals, using $K$ sub-carriers and $L$ symbols with a total symbol duration $T_\text{sym}=T+T_\text{CP}$, where $T=1/\Delta_f$ is the elementary symbol duration, $\Delta_f$ is the subcarrier spacing, and $T_\text{CP}$ is the \ac{CP} duration. The transmitted complex baseband signal is expressed as
\begin{equation}\label{eq_st} 
s(t)= \sqrt{\frac{P_\text{tot}}{K}}\sum_{\ell=0}^{L-1} \sum_{k=0}^{K-1} x_{k,\ell} e^{\jmath2\pi k\Delta_f (t-\ell T_\text{sym})} \operatorname{rect}\left(\frac{t-\ell T_\text{sym}}{T_\text{sym}}\right)\,,
\end{equation}
where $P_\text{tot}$ is the total transmitted power over all subcarriers, $x_{k,\ell}\in\mathbb{C}$ is the transmitted pilot symbol during the $\ell^{\text{th}}$ transmission on the $k^{\text{th}}$ sub-carrier with $\lvert x_{k,\ell}\rvert=1$, $\operatorname{rect}(t)=1$ when $t \in[0,1]$, and $0$ otherwise, and $t\in[0,LT_\text{sym}]$. The transmitted passband signal is expressed as $\tilde{s}(t)=\Re\{\text{exp}(\jmath2\pi f_c t) s(t)\}$. 

\subsubsection{Received Passband Signal at the UE}
The channel between the satellite and the \ac{UE} comprises two paths, a direct \ac{LoS} satellite-\ac{UE} path (`su') and a 
satellite-\ac{RIS}-\ac{UE} path (`sru', called the \ac{RIS} path), i.e., no multipath effect assumed.\footnote{ Attributes of these two paths, e.g., channels, delays, and Dopplers, will be sub-scripted by $i\in\{\text{su, sru}\}$, respectively. Additionally, the attributes of all paths between individual entities might be alternatively sub-scripted by $j\in\{\text{su, sr, and ru}\}$. Finally, parameters that pertain to the \ac{RIS} path only, e.g., \ac{AoA}, \ac{AoD}, and steering vectors, will be sub-scripted by $g\in\{\text{rs, ru}\}$. The first letter of $g$ is associated with the origin of the vector/center of the angle and the second letter with the destination.}
The received passband signal at the \ac{UE} in the time domain is $\tilde{y}(t)=\tilde{y}_\text{su}(t)+\tilde{y}_\text{sru}(t)+n(t)$, where $n(t)$ is the thermal noise at the receiver with \ac{PSD} $N_0$ and
\begin{align}
    \tilde{y}_\text{su}(t) &=\rho_\text{su}\tilde{s}(t-\tau_\text{su}+\nu_\text{su}t) \label{eq_ysu_pass}\\
    \tilde{y}_\text{sru}(t)\!&=\label{eq_ysru_pass}\!\rho_\text{sru}\!\sum_{n=0}^{N-1}\tilde{s}(t\!-\!\tau_\text{sru}\!+\!\nu_\text{sru}t\!-\!\tau_{n}(t)\!-\!\tau_{\boldsymbol{\theta}_\text{rs},n}\!-\!\tau_{\boldsymbol{\theta}_\text{ru},n}) \,, 
\end{align}
in which $\rho_i$, $\tau_i$, and $\nu_i$ are the channel amplitude (due to path loss and antenna patterns and atmospheric effects, e.g., tropospheric attenuation), the initial delay, and the frequency shift factor of the $i^{\text{th}}$ path $(i\in\{\text{su, sru}\})$, and $\tau_{n}(t)$ is the controlled delay of the $n^{\text{th}}$ RIS element at time $t$. Moreover,  $\tau_{\boldsymbol{\theta}_\text{rs},n}$ captures the delay/time-advance\footnote{A positive $\tau_{\boldsymbol{\theta},n}$ value corresponds to a delay with respect to the phase center while a negative value conveys a time advance.} of the impending signal at the 
$n^{\text{th}}$ RIS element (from the satellite), compared to the RIS phase center. Similarly, $\tau_{\boldsymbol{\theta}_\text{ru},n}$ is the delay at the $n^{\text{th}}$ RIS element of the signal towards the \ac{UE} compared to the RIS phase center. Under the plane-wave model, these two delays depend on the \ac{AoA} at the RIS, $\boldsymbol{\theta}_\text{rs}$, and the \ac{AoD} from the RIS, $\boldsymbol{\theta}_\text{ru}$, respectively. Correspondingly, $\boldsymbol{\theta}_\text{rs}\!=\!(\theta^\text{az}_\text{rs},\theta^\text{el}_\text{rs})$ and $\boldsymbol{\theta}_\text{ru}=(\theta^\text{az}_\text{ru},\theta^\text{el}_\text{ru})$ comprise the azimuth and elevation \ac{AoA} and \ac{AoD} tuples of the \ac{RIS}, respectively, in the \ac{RIS}'s \ac{LCS}. From simple geometry, it follows that $\tau_{\boldsymbol{\theta},n}=-\boldsymbol{p}^\top_n \boldsymbol{u}(\boldsymbol{\theta})/c$ is affected by the \ac{AoA}/\ac{AoD} tuple, $\boldsymbol{\theta}$, where $\boldsymbol{u}(\boldsymbol{\theta})=[\cos(\theta^\text{az})\cos(\theta^\text{el}), \sin(\theta^\text{az})\cos(\theta^\text{el}), \sin(\theta^\text{el})]^{\top}$.

\subsection{Geometric Relations}\label{Sec:GeometricRelations}
The path delays are computed as $ \tau_\text{su}=d_\text{su}/c+\delta$, $\tau_\text{sru}=\tau_\text{sr}+\tau_\text{ru}$, $\tau_\text{sr}=d_\text{sr}/c$, and $\tau_\text{ru}=d_\text{ru}/c+\delta$, where $d_j$ is the initial range of the $j^{\text{th}}$ path at $t=0$ ($j\in\{\text{su, sr, and ru}\}$). The frequency shifts in the given paths are modeled as $\nu_\text{su}={\boldsymbol{v}_\text{sat}^{\top}
(\boldsymbol{p}-\boldsymbol{p}_\text{sat})}/({d_\text{su}c}) + \delta_f$, $\nu_\text{sru}=\nu_\text{sr}+\nu_\text{ru}$, $\nu_\text{sr}=-{\boldsymbol{v}_\text{sat}^{\top} \boldsymbol{R}\boldsymbol{u}(\boldsymbol{\theta}_\text{rs})}/{c}$, and $\nu_\text{ru}=\delta_f$.  Finally, the \ac{RIS}'s azimuth and elevation \ac{AoA}/\ac{AoD} in the \ac{LCS} are computed as $\theta^\text{az}_g=\arctan([\boldsymbol{R}^\top\Delta\boldsymbol{p}_g]_2/[\boldsymbol{R}^\top\Delta\boldsymbol{p}_g]_1)$ and $\theta^\text{el}_g=\arcsin([\boldsymbol{R}^\top\Delta\boldsymbol{p}_g]_3/d_g)$, respectively. Here, $\Delta\boldsymbol{p}_g$ denotes the vector from RIS to satellite (for $g=\text{rs}$) or from RIS to UE (for $g=\text{ru}$) in the \ac{GCS}.

\subsection{Assumptions} \label{sec:assumptions}
In general, $\rho_i$, $\nu_i$, and $\boldsymbol{\theta}_{\text{rs}}$ are all functions of time~$t$, as they depend on the varying location of the \ac{LEO} satellite. However, we will operate under conditions (see Section \ref{sec:results}) under which the Doppler shift caused by the satellite is constant over $t$ as $\boldsymbol{v}_\text{sat}$ is assumed to be constant over that period and $\boldsymbol{\theta}_{\text{rs}}(LT_\text{sym}) - \boldsymbol{\theta}_{\text{rs}}(0)\approx0$. The time dependence notation was omitted for $\alpha_i$ to simplify the expressions. However, $\alpha_i$ changes over time in the generative model. We also assume that the \ac{RIS} and the \ac{UE} are in close vicinity, thus~$\nu_\text{su}\approx \nu_\text{sr}$. By further assuming $\delta_f\ll\nu_\text{sr}$, we obtain that $\nu_\text{su}\approx \nu_\text{sru}$.\footnote{These Doppler assumptions are utilized solely by the estimation algorithm and do not influence the generative model. Therefore, even if these assumptions are violated (e.g., the UE is mobile or located far from the RIS), the generative model remains accurate, though the estimation performance may degrade. The extent to which these assumptions hold and their impact on performance will be examined in future work.} We also assume that the RIS phase configurations change slowly, at a rate of one configuration per OFDM symbol.

\section{Discrete-time observation model}\label{sec: detailedModeling}
\subsection{Received Baseband Signal at the UE}
The baseband noise-free signals corresponding to the \ac{LoS} and \ac{RIS} paths after downconversion of \eqref{eq_ysu_pass}--\eqref{eq_ysru_pass} are
\begin{align}\label{demodulated}
    y_\text{su}(t)&=\alpha_\text{su} e^{\jmath2\pi f_c \nu_\text{su} t} s(t-\tau_\text{su}+\nu_\text{su}t)\\
\label{demodulatedRIS}
    y_\text{sru}(t)&=\alpha_\text{sru} e^{\jmath2\pi f_c \nu_\text{sru} t} s(t-\tau_\text{sru}+\nu_\text{sru}t) G_{\text{RIS}}(t)\,,
\end{align}
where $\alpha_i=\rho_i \exp(-\jmath 2 \pi f_c \tau_i)$, 
$G_{\text{RIS}}(t)=(\mathbf{a}(\boldsymbol{\theta}_\text{rs}) \odot \mathbf{a}(\boldsymbol{\theta}_\text{ru}))^{\top} \boldsymbol{\omega}(t)\in\mathbb{C}$ is the \ac{RIS} beamforming gain, and $\mathbf{a}(\boldsymbol{\theta})\in\mathbb{C}^{N}$ is the \ac{RIS}'s steering vector given the angle tuple $\boldsymbol{\theta}$. The $n^{\text{th}}$ element of $\mathbf{a}(\boldsymbol{\theta})$ is modeled as $\text{a}_{n}(\boldsymbol{\theta})=\exp(-\jmath2\pi f_c \tau_{\boldsymbol{\theta},n})$. Note here that the term $k\Delta_f\tau_{\boldsymbol{\theta},n}$ was omitted from $\text{a}_{n}(\boldsymbol{\theta})$ since $K\Delta_f\tau_{\boldsymbol{\theta},n}\ll1, \, \forall \boldsymbol{\theta}, n$.

Sampling (\ref{demodulated}) and (\ref{demodulatedRIS}) at $t=\ell T_\text{sym}+T_\text{CP}+\kappa T/K$, i.e., after \ac{CP} removal and stacking the samples per OFDM symbol along columns indexed by $\ell$,
results in
\begin{equation}\label{sampled}
\begin{split}
    Y_{\text{su},\kappa,\ell}= &\alpha_\text{su} e^{\jmath2\pi f_c \nu_\text{su}(\ell T_\text{sym}+T_\text{CP}+\kappa T/K)}\\
    &s((1+\nu_\text{su})(\ell T_\text{sym}+T_\text{CP}+\kappa T/K)-\tau_\text{su})
    \end{split}
\end{equation}
\begin{equation}\label{sampledRIS}
\begin{split}
    Y_{\text{sru},\kappa,\ell}= &\alpha_\text{sru} G_{\text{RIS},\ell} e^{\jmath2\pi f_c \nu_\text{sru}(\ell T_\text{sym}+T_\text{CP}+\kappa T/K)}\\
    &s((1+\nu_\text{sru})(\ell T_\text{sym}+T_\text{CP}+\kappa T/K)-\tau_\text{sru})\,,
    \end{split}
\end{equation}
where $\kappa=0,\dots,K-1$ and $G_{\text{RIS},\ell}$ is the sampled version of $G_{\text{RIS}}(t)$. Here, the constant phase shift caused by $f_c\nu_iT_\text{CP}$ in the exponent terms can be absorbed into $\alpha_i$ and the known phase shift caused by $k\Delta_fT_\text{CP}$ in $s(\cdot)$ can be merged to $x_{k,l}$ to simplify the expression. 

Several standard assumptions from terrestrial positioning can be followed while others cannot, as they are no longer valid in this setup due to high Doppler shifts caused by the \ac{LEO} satellite.\footnote{According to the simulation parameters shown in Table \ref{table}, the Doppler shift factor is $\approx 1.8*10^{-5}$. Hence, $K\Delta_f\nu_iT_\text{CP}\approx 2.6^{-3}\ll1$, $K\Delta_f\nu_iT\approx0.04\ll1$, $K\Delta_f \nu_iLT_\text{sym}\approx10.1>1$, $f_c\nu_iLT_\text{sym}\approx675\gg1$, and $f_c\nu_iT\approx2.5$.} For instance, we can neglect the terms $\nu_iT_\text{CP}$ and $\nu_i\kappa T/K$ in $s(\cdot)$, as $K\Delta_f\nu_iT_\text{CP}\ll1$ and $K\Delta_f\nu_iT\ll1$, respectively. In contrast, the $\nu_i\ell T_\text{sym}$ term in $s(\cdot)$ cannot be ignored, as $K\Delta_f \nu_iLT_\text{sym}>1$, i.e., the effect of large time-bandwidth product $K \Delta_f L T_\text{sym} $ (also called intersubcarrier Doppler effect \cite{ofdm_isac_2020}). Likewise, the carrier frequency phase shift caused by $f_c\nu_i\ell T_\text{sym}$ and $f_c\nu_i\kappa T/K$ cannot be ignored, as $f_c\nu_iLT_\text{sym}\gg1$ and $f_c\nu_iT\approx1$. The latter two effects are called the \emph{slow-time and fast-time Doppler effects}, respectively. Hence, (\ref{sampled}) and (\ref{sampledRIS}) can be simplified to be
\begin{align}\label{simplified}
\begin{split}
    Y_{\text{su},\kappa,\ell}= & \alpha_\text{su} e^{\jmath2\pi f_c \nu_\text{su}(\ell T_\text{sym}+\kappa T/K)}\\
    & s(\ell T_\text{sym}+\kappa T/K-\tau_\text{su}+\nu_\text{su}\ell T_\text{sym})
    \end{split}\\
\label{simplifiedRIS}
\begin{split}
    Y_{\text{sru},\kappa,\ell}= & \alpha_\text{sru} G_{\text{RIS},\ell} e^{\jmath2\pi f_c \nu_\text{sru}(\ell T_\text{sym}+\kappa T/K)}\\
    & s(\ell T_\text{sym}+\kappa T/K-\tau_\text{sru}+\nu_\text{sru}\ell T_\text{sym})\,.
    \end{split}
\end{align}

\subsection{Observations in Compact Matrix Form}
By inserting \eqref{eq_st}, the simplified \ac{LoS} and \ac{RIS} path signals in \eqref{simplified}--\eqref{simplifiedRIS} can be re-written in matrix form, respectively, as
\begin{align} \label{matrixForm}
    \boldsymbol{Y}_{\text{su}} &=\alpha_\text{su}\boldsymbol{A}_\text{su}\odot\boldsymbol{F}^{\mathsf{H}}\left(\boldsymbol{B}_\text{su}\odot\boldsymbol{C}_\text{su}\odot\boldsymbol{X}\right)\\
\label{matrixForm_sru}    \boldsymbol{Y}_{\text{sru}} &=\alpha_\text{sru}\boldsymbol{A}_\text{sru}\odot\boldsymbol{F}^{\mathsf{H}}\left(\boldsymbol{B}_\text{sru}\odot\boldsymbol{C}_\text{sru}\odot\boldsymbol{G_{\text{RIS}}}\odot\boldsymbol{X}\right)\,,
\end{align}
where $\boldsymbol{X} \in \mathbb{C}^{K\times L}$ incorporates the transmitted pilot symbols and $\sqrt{{P_\text{tot}}/{K}}$, $\boldsymbol{F}\in \mathbb{C}^{K\times K}$ is the unitary FFT matrix, $\boldsymbol{A}_i \in \mathbb{C}^{K\times L}$ and $\boldsymbol{C}_i \in \mathbb{C}^{K\times L}$ capture the fast-time and slow-time Doppler effects, respectively, of the $i^{\text{th}}$ path, $\boldsymbol{B}_i \in \mathbb{C}^{K\times L}$ captures the sub-carrier phase shift and Doppler effects of the $i^{\text{th}}$ path, and $\boldsymbol{G_{\text{RIS}}}\in \mathbb{C}^{K\times L}$ encapsulates the \ac{RIS} response. The elements of the matrices above are as follows  ${A}_{i,\kappa,\ell}=e^{\jmath 2 \pi f_c \nu_i \kappa/K T}$, ${B}_{i,k,\ell}= e^{-\jmath 2 \pi k \Delta_f (\tau_i-\nu_i \ell T_\text{sym})}$, ${C}_{i,k,\ell}=e^{\jmath 2 \pi f_c \nu_i \ell T_\text{sym}}$, and ${G_{\text{RIS}}}_{k,\ell}=G_{\text{RIS},\ell}$. 
Finally, combining \eqref{matrixForm} and \eqref{matrixForm_sru}, the received signal in the time domain can be expressed as
\begin{align}\label{ch-model-time}
\boldsymbol{Y}&=\sum_{i\in \{\text{su},\text{sru}\}} \boldsymbol{A}_i\odot \boldsymbol{F}^{\mathsf{H}} (\boldsymbol{H}_i\odot \boldsymbol{X})+\boldsymbol{N}\,,
\end{align}
where $\boldsymbol{H}_i\in \mathbb{C}^{K\times L}$ encapsulates the effects of $\alpha_i$, $\boldsymbol{B}_i$, $\boldsymbol{C}_i$, and $\boldsymbol{G_{\text{RIS}}}$ (for the \ac{RIS} path),  $\boldsymbol{N} \sim \mathcal{CN}(\boldsymbol{0}_{K \times L},\sigma^2\boldsymbol{I}_{K \times L})$, $\sigma^2=N_0N_f$ is the noise variance at the \ac{UE}, and $N_f$ is the \ac{NF} of the \ac{UE}.

\section{Localization Method}
\subsection{Channel Parameter Estimation}
The unknown channel parameters in \eqref{ch-model-time} comprise $\alpha_\text{su}, \alpha_\text{sru}, \tau_\text{su}, \tau_\text{sru},\nu_\text{su}, \nu_\text{sru}$, and $\boldsymbol{\theta}_\text{ru}$. To avoid using a 6D \ac{MLE}\footnote{Here, 6D \ac{MLE} can be used instead of 8D \ac{MLE} because $\alpha_\text{su}$ and $\alpha_\text{sru}$ can be estimated in closed-form as a function of the remaining unknowns.}, we utilize the fact that the \ac{RIS} path is much weaker than the \ac{LoS} path, i.e., $\left\lvert\alpha_\text{su} \right\lvert \gg \left\lvert\alpha_\text{sru}\right\lvert$. Hence, we treat $\boldsymbol{Y}$ as if it only constitutes the effects of the dominant \ac{LoS} path, and proceed with estimating the \ac{LoS} channel parameters $(\alpha_\text{su}, \tau_\text{su}, \nu_\text{su})$. Then, we can subtract the reconstructed \ac{LoS} path from $\boldsymbol{Y}$ and estimate the \ac{RIS} path parameters. 

\begin{remark}[Range and Doppler ambiguities]
    The products $\Delta_f\tau_i$ and $f_c\nu_i T_\text{sym}$ can be $>1$, due to the long transmission distance and the high speed of the \ac{LEO} satellite, respectively. Hence, those products can be re-written as $N_{\tau_i}+\Delta_f\tilde{\tau}_i$ and $N_{\nu_i}+f_c\tilde{\nu}_iT_\text{sym}$, respectively, where $N_{\tau_i}=\lfloor\Delta_f\tau_i\rfloor\in \mathbb{N}$, $N_{\nu_i}=\lfloor f_c\nu_i T_\text{sym}\rfloor\in \mathbb{N}$, $\tilde{\tau}_i=\tau_i-N_{\tau_i}/\Delta_f$, and $\tilde{\nu}_i=\nu_i-N_{\nu_i}/(f_cT_\text{sym})$. Since the UE and RIS are in close proximity, the integer values of the LoS and RIS path delay will mostly be identical, as will those for the Doppler. However, to avoid possible ambiguities, differential delays and Dopplers will be exploited instead of absolute ones so that integer ambiguities will cancel out.
\end{remark}

\subsubsection{LOS Path Channel Parameter Estimation}
First, we eliminate the fast-time Doppler effect, $\boldsymbol{A}_\text{su}$, and the sub-carrier Doppler shift (the second exponential term in $\boldsymbol{B}_\text{su}$, which depends on $\nu_\text{su}$) from \eqref{ch-model-time}, as they will negatively affect the delay estimation. To do that, we rely on the assumptions from Sec.~\ref{sec:assumptions} to coarsely set $\hat{\nu}_\text{su}\approx\nu_\text{sr}$. This allows us to eliminate the bulk of the Doppler effects by computing
\begin{align}
    \boldsymbol{\mathring{Y}}&=\boldsymbol{F} (\boldsymbol{\hat{A}}^*_\text{su} \odot \boldsymbol{Y}) \label{Y-Circle}\\
    \boldsymbol{\breve{Y}}&= \boldsymbol{\widehat{B}}^*_{\hat{\nu}_\text{su}} \odot \boldsymbol{\mathring{Y}}\,, \label{Y-HalfCircle}
\end{align}
where $\boldsymbol{\hat{A}}^*_\text{su}$ and $\boldsymbol{\widehat{B}}^*_{\hat{\nu}_\text{su}}$ are the complex conjugates of the estimates of $\boldsymbol{A}_\text{su}$ and $\boldsymbol{B}_{\nu_\text{su}}$, respectively, with elements  $\hat{A}^*_{\text{su},\kappa,\ell}=e^{-\jmath 2 \pi f_c \hat{\nu}_\text{su} \kappa/K T}$ and ${\widehat{B}}^*_{\hat{\nu}_\text{su}k,\ell}= e^{-\jmath 2 \pi k \Delta_f \hat{\nu}_\text{su} \ell T_\text{sym}}$, respectively. Then, absorbing constant  $\boldsymbol{X}$ into $\alpha_\text{su}$ and introducing ${B}^{\text{res}}_{\text{su},k,\ell}=e^{-\jmath 2 \pi k \Delta_f \tau_i}$, we find
\begin{align}
    \boldsymbol{\breve{Y}} & \approx \alpha_\text{su}\left(\boldsymbol{B}^{\text{res}}_\text{su}\odot\boldsymbol{C}_\text{su}\right) + \boldsymbol{N}\\
    & = \alpha_\text{su} \boldsymbol{b}(\tau_\text{su})\boldsymbol{c}^\top(\nu_\text{su}) + \boldsymbol{N}, \label{eq:outer-product}
\end{align}
where $b_{k}(\tau_{\text{su}})=e^{-\jmath 2 \pi k \Delta_f \tau_{\text{su}}}$ and $c_{\ell}( \nu_{\text{su}})=e^{\jmath 2 \pi f_c \nu_{\text{su}} \ell T_\text{sym}}$, since all columns in  $\boldsymbol{B}^{\text{res}}_\text{su}$ are identical, and all rows in $\boldsymbol{C}_\text{su}$ are also identical. From  \eqref{eq:outer-product}, the delay and Doppler can directly be estimated via a 2D-FFT or by 2 separate 1D-FFT with non-coherent integration. Under 2D-FFT, 
\begin{align} \label{17}
    [\hat\tau_{\text{su}},\hat\nu_{\text{su}}]=\arg \max_{\tau,\nu} \lvert\boldsymbol{b}^{\mathsf{H}}(\tau) \boldsymbol{\breve{Y}} \boldsymbol{c}^*(\nu)\rvert\,.
\end{align}

To refine these initial estimates $\hat{\tau}_\text{su}$ and $\hat{\nu}_\text{su}$, we perform an iterative 2D \ac{MLE} search with $M$ iterations. Each iteration halves the search space of the previous iteration. The \ac{MLE} at each iteration can be expressed as
\begin{equation}\label{18}
    [\hat{\tau}_\text{su},\hat{\nu}_\text{su}]= \operatorname*{argmin}_{\tau_\text{su},\nu_\text{su}} \lVert \boldsymbol{\Pi}_{\boldsymbol{z}}^{\perp}\boldsymbol{\mathring{y}}\rVert\,,
\end{equation}
where $\boldsymbol{\mathring{y}}=\operatorname{vec}(\boldsymbol{\mathring{Y}})$, $\boldsymbol{\Pi}_{\boldsymbol{z}}^{\perp} \triangleq\textbf{I}-\boldsymbol{\Pi}_{\boldsymbol{z}}$, $\boldsymbol{\Pi}_{\boldsymbol{z}} \triangleq \boldsymbol{z}\left(\boldsymbol{z}^{\mathsf{H}} \boldsymbol{z}\right)^{-1} \boldsymbol{z}^{\mathsf{H}}$ denotes the orthogonal projector onto the column space of $\boldsymbol{z}$, $\boldsymbol{z}=\operatorname{vec}(\boldsymbol{Z})$, $\boldsymbol{Z}=\boldsymbol{\hat{B}}_\text{su} \odot \boldsymbol{\hat{C}}_\text{su} \odot \boldsymbol{X}$, and $\boldsymbol{\hat{B}}_\text{su}$ and $\boldsymbol{\hat{C}}_\text{su}$ are the estimated versions of $\boldsymbol{B}_\text{su}$ and $\boldsymbol{C}_\text{su}$, respectively, using $\hat{\tau}_\text{su}$ and $\hat{\nu}_\text{su}$. Finally, $\hat{\alpha}_\text{su}$ is estimated as follows $\hat{\alpha}_\text{su}= \boldsymbol{z}^{\mathsf{H}}\boldsymbol{\mathring{y}}/\lVert\boldsymbol{z}\rVert^2$.

\subsubsection{RIS-path channel parameter estimation} \label{sec:RIS-path-estimation}
The first step toward estimating the \ac{RIS}-path channel parameters is to eliminate the \ac{LoS} path component from (\ref{ch-model-time}), using the estimated \ac{LoS} channel parameters to form $\boldsymbol{\hat{Y}}_\text{sru}=\boldsymbol{Y}-\boldsymbol{\hat{Y}}_\text{su}$. Then, we utilize the assumptions highlighted in Sec. \ref{sec:assumptions} to set $\hat{\nu}_\text{sru}=\hat\nu_\text{su}$, which enable us to speed up the estimation process of the other parameters by decoupling the \ac{AoD} and Doppler effects in the slow-time dimension at the expense of biasing the \ac{RIS} path Doppler estimator. Next, we eliminate the fast-time Doppler effect and the sub-carrier Doppler shift effect in a similar fashion to (\ref{Y-Circle})--(\ref{Y-HalfCircle}). Similarly, we eliminate the slow-time Doppler shift effect by using $\boldsymbol{\hat{C}}_\text{sru}$ to form $\boldsymbol{\breve{Y}}_\text{sru}$, where $\boldsymbol{\hat{C}}_\text{sru}$ is the estimated version of $\boldsymbol{C}_\text{sru}$, by utilizing $\hat{\nu}_\text{sru}$, and $\boldsymbol{\breve{Y}}_\text{sru}\in \mathbb{C}^{K\times L}$ is the Doppler-free version of $\boldsymbol{\hat{Y}}_\text{sru}$, with (again asborbing the constant $\boldsymbol{X}$ into $\alpha_\text{sru}$)
\begin{align}
    \boldsymbol{\hat{Y}}_\text{sru} &  \approx \alpha_\text{sru}\left(\boldsymbol{B}_\text{sru}\odot\boldsymbol{G_{\text{RIS}}}\right) + \boldsymbol{N}\\
     & = \alpha_\text{sru} \boldsymbol{b}(\tau_{\text{sru}}) \boldsymbol{g}^\top_{\text{RIS}} + \boldsymbol{N}\,,
\end{align}
where $\boldsymbol{g}_{\text{RIS}}=[G_{\text{RIS},0},\ldots, G_{\text{RIS},L-1}]^\top$ is the \ac{RIS} response.
Here, we used the fact that the
sub-carrier dimension (the rows of $\boldsymbol{\hat{Y}}_\text{sru}$) is solely modulated by the \ac{RIS}-path delay $\tau_\text{sru}$, and the slow-time dimension is modulated solely by the \ac{RIS} response $\boldsymbol{g}_{\text{RIS}}$, which is affected by the unknown \ac{AoD} $\boldsymbol{\theta}_\text{ru}$. The RIS path delay can be estimated via a 1D-FFT and non-coherent integration over slow time
\begin{align}\label{21}
    \hat{\tau}_{\text{sru}}=\arg \max_{\tau} \sum_{\ell=0}^{L-1} | \boldsymbol{b}^{\mathsf{H}}(\tau) \boldsymbol{\hat{y}}_{\text{sru},\ell}|\,,
\end{align}
where $\boldsymbol{\hat{y}}_{\text{sru},\ell}$ denotes the $\ell$-th column of $\boldsymbol{\hat{Y}}_\text{sru}$. Next, we perform an iterative 2D \ac{MLE} grid search on the \ac{AoD} tuple, followed by an iterative 1D \ac{MLE} grid search on the \ac{RIS}-path delay. In each step, we utilize initial estimates to perform coherent integration. The 2D \ac{AoD} \ac{MLE} search at each iteration is as
\begin{equation}\label{22}
[\hat{\theta}^\text{az}_\text{ru},\hat{\theta}^\text{el}_\text{ru}]=\operatorname*{argmin}_{\theta^\text{az}_\text{ru},\theta^\text{el}_\text{ru}}\lVert \boldsymbol{\Pi}_{\boldsymbol{z}_\text{AoD}}^{\perp}\boldsymbol{\breve{y}}_\text{AoD}\rVert\,,
\end{equation}
where $\boldsymbol{\breve{y}}_\text{AoD}=\boldsymbol{\breve{Y}}^\top_\text{sru} \boldsymbol{b}^*(\hat{\tau}_\text{sru})\in\mathbb{C}^L$ is the coherent integration of $\boldsymbol{\breve{Y}}_\text{sru}$ over sub-carriers and $\boldsymbol{z}_\text{AoD}=\boldsymbol{g}_{\text{RIS}}$, which is a function of the  $\boldsymbol{\theta}_\text{ru}$ tuple. Similarly, the 1D \ac{MLE} search formulation  at each iteration is as follows 
\begin{equation}\label{23}
    \hat{\tau}_\text{ru}= \operatorname*{argmin}_{\tau_\text{sru}} \lVert \boldsymbol{\Pi}_{\boldsymbol{z}_\tau}^{\perp}\boldsymbol{\breve{y}}_\tau\rVert-\tau_\text{sr}\,,
\end{equation}
where $\boldsymbol{\breve{y}}_\tau=\boldsymbol{\breve{Y}}_\text{sru}\boldsymbol{\hat{g}}_{\text{RIS}}\in\mathbb{C}^{K}$ is the coherent integration of $\boldsymbol{\breve{Y}}_\text{sru}$ over slow-time, $\boldsymbol{\hat{g}}_{\text{RIS}}$ is the estimated version of $\boldsymbol{g}_{\text{RIS}}$, and $\boldsymbol{z}_\tau=\boldsymbol{b}(\hat{\tau}_\text{sru})$. Finally, the estimate of the \ac{RIS}-path channel gain is as follows $\hat{\alpha}_\text{sru}= \boldsymbol{z}^{\mathsf{H}}_\tau\boldsymbol{\breve{y}}_\tau/\lVert\boldsymbol{z}_\tau\rVert^2$.

\subsection{Position, Time-bias, and CFO estimation}
The position of the \ac{UE} can be estimated as $\hat{\boldsymbol{p}}=\boldsymbol{p}_\text{RIS}+\hat{d}_\text{ru}\boldsymbol{R}\boldsymbol{u}(\boldsymbol{\hat{\theta}}_\text{ru})$, where $\hat{d}_\text{ru}$ is the estimated range between the \ac{RIS} and the \ac{UE}. To estimate $\hat{d}_\text{ru}$, we solve the following minimization problem
\begin{equation}
\begin{aligned}
 \hat{d}_\text{ru} = \operatorname*{argmin}_{d_\text{ru}} (
   (&d_\text{sr}+d_\text{ru})-(\hat{\tau}_\text{sru}-\hat{\tau}_\text{su})c\\
   &-\lVert \boldsymbol{p}_\text{sat} -\boldsymbol{p}_\text{RIS}-d_\text{ru}\boldsymbol{R}\boldsymbol{u}(\boldsymbol{\hat{\theta}}_\text{ru})\rVert)^2\,.
\end{aligned}
\end{equation}
Finally, the time bias and \ac{CFO} are estimated as follows $\hat{\delta}= \hat{\tau}_\text{su}-\hat{d}_\text{su}/c$, $\hat{\delta}_f= \hat{\nu}_\text{su}-\boldsymbol{v}_\text{sat}^{\top}\cdot \boldsymbol{u}(\boldsymbol{\hat{\theta}}_\text{su})/c$. The overall localization methodology is summarized in Alg. \ref{Alg}.

\begin{algorithm}
\caption{Proposed Localization Algorithm}\label{Alg}
\begin{algorithmic}[1]
    \State \textbf{Input:} Received signal $\boldsymbol{Y}$ in the time domain.
        \State Compensate fast-time Doppler with $\hat{\nu}_\text{su}\approx \nu_\text{sr}$  \eqref{Y-Circle}-\eqref{Y-HalfCircle}.
        \State \textbf{2D FFT Search:} Coarse estimate of $\hat{\tau}_\text{su}$ and $\hat{\nu}_\text{su}$ \eqref{17}.
        \State \textbf{2D MLE Search:} Fine estimate of $\hat{\tau}_\text{su}$ and $\hat{\nu}_\text{su}$ \eqref{18}.
        \State Estimate $\hat{\alpha}_\text{su}$ and determine residual $\boldsymbol{\hat{Y}}_\text{sru}$. 
        \State Compensate the RIS path Doppler with $\hat{\nu}_\text{sru}\approx\hat{\nu}_\text{su}$.
        \State \textbf{1D FFT Search:} Coarse estimate of $\hat{\tau}_\text{ru}$ \eqref{21}.
        \State \textbf{2D MLE Search:} Fine estimate of $\boldsymbol{\hat{\theta}}_\text{ru}$ \eqref{22}.
        \State \textbf{1D MLE Search:} Fine estimate of $\hat{\tau}_\text{ru}$ \eqref{23}.
        \State \textbf{Output:} UE position $\boldsymbol{\hat{p}}$, clock bias $\hat{\delta}$, and CFO $\hat{\delta}_f$
\end{algorithmic}
\end{algorithm}

\subsection{Complexity Analysis}
For the LoS path, the complexity stems from both the 2D-FFT and the 2D \ac{MLE} refinement. The 2D-FFT scales as $\mathcal{O}(N_K N_L \log_2(N_K N_L))$, where $N_K$ and $N_L$ represent the FFT size in delay and Doppler dimension. The \ac{MLE} refinement for the LoS parameters has complexity $\mathcal{O}(N^2_g MK L)$, where $N_g$ is the number of grid points per dimension. For the RIS path, the 1D FFT delay estimation has complexity $\mathcal{O}(L N_K \log N_K)$. The 2D-AoD and the 1D delay \ac{MLE} searches have complexities of $\mathcal{O}(N^2_g M L)$ and $\mathcal{O}(N_g M K)$, respectively. 
On the other hand, a 6D \ac{MLE} grid search has a complexity of $O(N_g^6MKL)$. Hence, the proposed method drastically reduces the complexity.

\section{Simulation Setup and Results}

\subsection{Simulation Setup}
The simulation parameters, including the transmission power, atmospheric losses, antenna gains, etc, were guided by 3GPP's R-16 technical report (TR.38.821) (scenario C1, set-1, case 9) \cite{3GPP38821}. The \ac{RIS} element pattern was modeled according to \cite{ellingson_path_2021}.
The simulation setup fixes all parameters, including the position of the \ac{UE}, the \ac{RIS}, and the \ac{LEO} satellite settings,\footnote{Fixing the satellite settings refers to its initial position and velocity. However, as highlighted in Sec. \ref{sec:assumptions}, the satellite's position will change over time, and hence the channel gains will also change accordingly.} and sweeps the transmission power of the \ac{LEO} satellite, i.e., $P_\text{tot} \text{[dBm]}=P_\text{3GPP}\text{[dBm]}+P_\text{sweep}\text{[dB]}$, where $P_\text{3GPP}$ is 3GPP's nominal satellite transmission power, highlighted in Table \ref{table}, and $P_\text{sweep}\in[-40,-35,\dots, -10]$ dB. For each corresponding \ac{SNR}, defined as  $\text{SNR}=P_\text{tot}T(\lvert\alpha_\text{su}\rvert^2L+\lvert\alpha_\text{sru}\rvert^2\lVert\boldsymbol{g}_\text{RIS}\rVert_2^2)/\sigma^2$, the \ac{RMSE} of the parameters is computed based on 100 Monte Carlo runs and compared to the corresponding \ac{CRLB}.\footnote{Following a similar methodology to the one shown in \cite{wang2024beamforming} while using the extended signal model highlighted in Sec. \ref{sec: detailedModeling}.} Table \ref{table} shows the simulation parameters used.

\begin{table}
\caption{Simulation parameters}
    \centering
\begin{tabular}{lll}
\hline \hline Parameter & Symbol & Value \\
\hline
\ac{UE} position & $\boldsymbol{p}$ & $[0, 10, 1.5]$ m\\
\ac{RIS} position & $\boldsymbol{p}_\text{RIS}$ & $[0, 0, 10]$ m\\
\ac{RIS} orientation & $\boldsymbol{R}$ & $\boldsymbol{I}_{3\times3}$ \\
\ac{RIS} dimensions & $N_x\times N_z$ & $10\times10$ \\
\ac{LEO} satellite orbital altitude & $h$ & $600$ km \\
\ac{LEO} satellite angles & $(\theta^\text{az}_\text{sat}, \theta^\text{el}_\text{sat})$ & $(90^\circ, 45^\circ)$ \\
Carrier frequency & $f_c$ & $2$ GHz\\
\# of sub-carriers & $K$ & $2000$ \\
\# of symbols & $L$ & $256$ \\
Sub-carrier spacing & $\Delta_f$ & $15$ kHz \\
Cyclic prefix duration & $T_\text{CP}$ & $7\%T$ \\
Time bias & $\delta$ & $1$ ns \\
\ac{CFO} factor & $\delta_f$ & $10^{-6}$ \\
Nominal 3GPP transmission power & $P_\text{3GPP}$ & $54$ dBm \\
Noise \ac{PSD} & $N_0$ & $-174$ dBm/Hz \\
Noise figure & $N_f$ & $7$ dB \\
IFFT/FFT size & $N_K,N_L$ & $2^{13}$ \\
\ac{MLE} grid size & $N_g$ & $10$ \\
\ac{MLE} \# of iterations & $M$ & $10$ \\
\ac{MLE} delay grid boundaries  & $\tau_\text{max}$& $\pm5/c$ s\\
\ac{MLE} Doppler grid boundaries & $\nu_\text{max}$& $\pm7/c$\\
\ac{MLE} \ac{AoD} grid boundaries  & $\theta_\text{max}$& $\pm45^\circ$\\
Location uncertainty for beamforming & $\sqrt{\text{diag}(\boldsymbol{\Sigma}_\text{p})}$& [1,1,1] m\\
\hline \hline
\end{tabular}
\label{table}
\end{table}

\subsubsection*{Satellite orbit}\label{sec:orbit}
In order to simulate appropriate range and Doppler measurements, we developed a simple satellite orbit function that takes the satellite's orbital altitude $h$ and the initial ascension (azimuth)-elevation angle tuple $\boldsymbol{\theta}_\text{sat}=(\theta^\text{az}_\text{sat}, \theta^\text{el}_\text{sat})$ as inputs, and provides the satellite position and velocity vectors as outputs. The computation of the satellite position at time $t$ is modeled as follows $\boldsymbol{p}_\text{sat} (t) = d_\text{sat}(t) \boldsymbol{u}(\boldsymbol{\theta}_\text{sat}(t))$, where $d_\text{sat}(t)$ is the distance from the center of the \ac{GCS} on the surface of the earth to the satellite at time $t$ and is computed as follows $d_\text{sat}(t)=R\sin(-\theta^\text{el}_\text{sat}(t))+\sqrt{R^2\sin^2(-\theta^\text{el}_\text{sat}(t))+2Rh+h^2}$, where $R$ is the radius of the earth. The velocity of the satellite at time $t$ is modeled as follows $\boldsymbol{v}_\text{sat} (t) = \dot d_\text{sat}(t)\boldsymbol{u}(\boldsymbol{\theta}_\text{sat}(t))+d_\text{sat} (t) \boldsymbol{\dot u}(\boldsymbol{\theta}_\text{sat}(t))$, where $\dot d_\text{sat}(t)$ and $\boldsymbol{\dot u}(\boldsymbol{\theta}_\text{sat}(t))$ are the time derivatives of $d_\text{sat}(t)$ and $\boldsymbol{u}(\boldsymbol{\theta}_\text{sat}(t))$, respectively. 

\subsubsection*{RIS configurations}\label{sec:RIS Config}
Two \ac{RIS} configuration strategies were tested in this work. In the first configuration strategy, the components of $\boldsymbol{\omega}_\ell$ were chosen randomly from $[0,2\pi)$. The second strategy assumes that the \ac{RIS} is roughly aware of the \ac{UE}'s location and beamforms the reflected signal toward that location in a stochastic fashion, i.e., the RIS beamforms at $\tilde{\boldsymbol{p}}\sim\mathcal{N}(\boldsymbol{p},\boldsymbol{\Sigma}_\text{p})$, where $\boldsymbol{\Sigma}_\text{p}\in\mathbb{R}^{3\times 3}$ is the covariance of the \ac{UE}'s location. The \ac{RIS} configuration for the $\ell^{\text{th}}$ transmission is computed as $\boldsymbol{\omega}_\ell=(\mathbf{a}(\boldsymbol{\theta}_\text{rs})\odot\mathbf{a}(\boldsymbol{\tilde{\theta}}_{\text{ru},\ell}))^*$, where $\boldsymbol{\tilde{\theta}}_{\text{ru},\ell}$ is the \ac{RIS}'s \ac{AoD} tuple towards the sampled \ac{UE} location, $\tilde{\boldsymbol{p}}_\ell$.

\subsection{Results and Discussion}\label{sec:results}
\begin{figure*}
\centering
\input{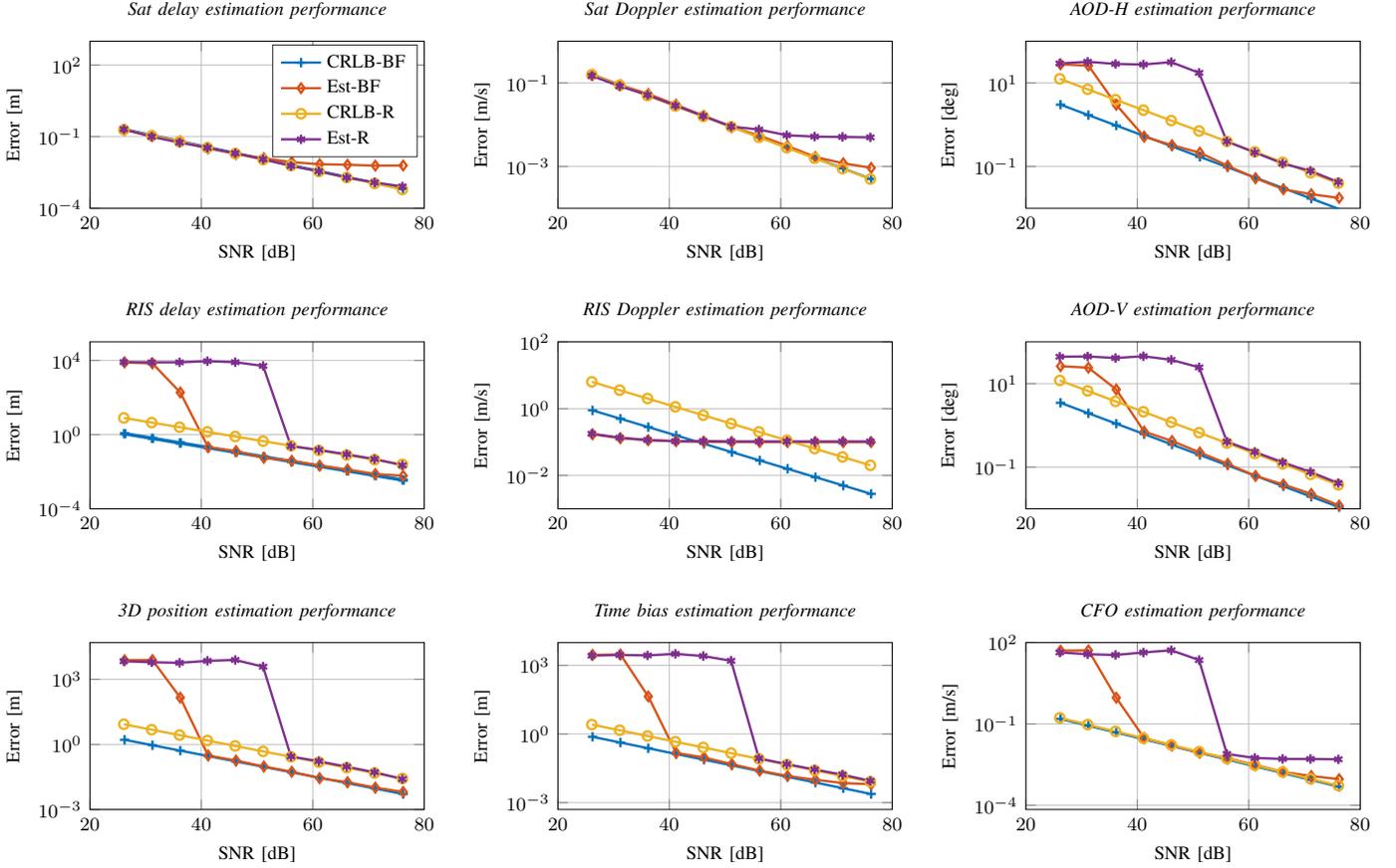}
\caption{Estimation performance vs CRLB with beamforming (BF) and random RIS configurations (R). The RIS-path Doppler estimator broke the CRLB bound as it is a biased estimator that is tested in favourable conditions (further discussions in Sec. \ref{sec:results}).}
\label{Fig: performance}
\vspace{-5mm}
\end{figure*}

The \ac{CRLB} and performance of the proposed estimator for the \ac{BF} and random configuration cases across multiple \ac{SNR} values are shown in Fig.~\ref{Fig: performance}. In general, it can be seen that the estimator can achieve the \ac{CRLB} performance at high \acp{SNR}. Moreover, we observe that, for most channel parameters, the estimator usually attains the bounds at lower transmission powers while using \ac{BF} compared to its random counterpart. The theoretical error bounds of these parameters are also lower when using \ac{BF}, compared to random \ac{RIS} configurations. This is mainly because the \ac{RIS} path's \ac{SNR} is $\approx20$ dB higher when utilizing \ac{BF}, compared to when using random configurations. Yet, not all parameters are affected the same by \ac{RIS} configuration. For instance, the estimation of the \ac{LoS} delay is enhanced slightly at higher SNRs while using random configurations. This is evident because the \ac{RIS}-path signal is considered as interference while estimating the \ac{LoS} parameters. On the contrary, the \ac{RIS}-path delay and \ac{AoD} estimation are enhanced by \ac{BF}. Likewise, the position estimate performance mirrors the performance of the estimation of the \ac{RIS}-path delay and \ac{AoD}. This means that the system is bottlenecked by the performance of the \ac{RIS}-path estimation. Hence, any further research should focus on enhancing the \ac{RIS}-path estimation performance, e.g., using different types of RIS like STAR-RIS and active RIS or enhancing the estimation algorithm. The same can also be said about the time-bias estimation, as it heavily relies on the quality of the position estimate. Additionally, it can be seen that there is a slight performance gap between the \acp{CRLB} of the \ac{BF} and random configuration scenarios for position and time bias estimation, whereas \ac{CFO} estimation does not have such a gap. This is because, unlike position and time bias estimation, \ac{CFO} estimation is more dependent on the quality of the \ac{LoS} Doppler estimation in comparison to the quality of the \ac{RIS}-path's parameter estimation. This is mainly because of how close the UE is to the RIS and the ability to use the known satellite-\ac{RIS} Doppler to extract the \ac{CFO} from the \ac{LoS} path. Finally, it is worth noting here the fact that the \ac{RIS}-path Doppler estimator breaks the bound in both scenarios. This is  caused by the usage of the biased estimator $\hat{\nu}_\text{sru}=\hat{\nu}_\text{su}$ in Sec.~\ref{sec:RIS-path-estimation}.

\subsection{Open Challenges and Future Research Directions}
As this work marks the first step toward exploring the 6G single-LEO single-RIS problem, many open challenges and extensions remain to be investigated. For instance, the system model can be extended to be more realistic by (i) adding uncertainty about the position and velocity of the satellite due to gravitational forces and atmospheric drag; (ii) adding uncertainty about the position and orientation of the RIS due to installation/calibration errors; (iii) proper modeling of time-varying atmospheric effects like the tropospheric, ionospheric, and scintillation effects; (iv) considering a more realistic orbit for the LEO satellite; (v) adding mobility to the UE based on typical vehicular mobility models/patterns; (vi) adding additional paths (multipath effects); (vii) testing under non-line-of-sight (NLoS) conditions; and (viii) adding other hardware impairments (e.g., phase noise, power amplifier non-linearities, RIS pixel failure and mutual coupling, etc.). 


\section{Conclusion}
In this paper, we showcased that a 6G single  \ac{LEO} satellite has the potential to provide localization services for ground users with the aid of a single \ac{RIS}. We derived a comprehensive channel model that accounts for the satellite's mobility, fast- and slow-time Doppler effects, atmospheric effects, time-bias, and \ac{CFO}. Via a novel low-complexity estimator, we showed that meter-level positioning error can be theoretically achieved at $30$ dB \ac{SNR}. The solution estimates the \ac{ToA} and Doppler of the two paths as well as the $\ac{AoD}$ of the \ac{RIS} path to estimate the \ac{UE}'s position, time-bias, and \ac{CFO}. It was shown that the proposed estimator can attain the \ac{CRLB} at high \ac{SNR} and that it is bottle-necked by the \ac{RIS} path's parameter estimation errors. Moreover, we showed that configuring the \ac{RIS} elements to stochastically beamform towards the user's area can significantly enhance the estimator's performance. Finally, we presented a list of open challenges and possible future research directions.

\balance 
\bibliographystyle{IEEEtran}
\bibliography{References,referencesZ}
\end{document}